\documentclass[conference]{IEEEtran}

\usepackage{cite}

\usepackage{mathptmx}
\usepackage{times}
\usepackage{graphicx}
\usepackage{amsmath}
\usepackage{amssymb}
\usepackage[T1]{fontenc}
\usepackage{enumerate}
\usepackage{color}
\usepackage{times}
\usepackage{wrapfig}

\begin{document}
\title{Occupational Fraud Detection Through Visualization}

\author{\IEEEauthorblockN{Evmorfia~N.~Argyriou, Aikaterini~A.~Sotiraki, Antonios~Symvonis}
\IEEEauthorblockA{Department of Mathematics, School of Applied Mathematical \& Physical Sciences \\
National Technical University of Athens, Greece \\
$\{$fargyriou,sotiraki,symvonis$\}$@math.ntua.gr } }


%Telephone: (800) 555--1212, Fax: (888) 555--1212}


\maketitle

\begin{abstract}
%\boldmath
Occupational fraud affects many companies worldwide causing them
economic loss and liability issues towards their customers and other
involved entities. Detecting internal fraud in a company requires
significant effort and, unfortunately cannot be entirely prevented.
The internal auditors have to process a huge amount of data produced
by diverse systems, which are in most cases in textual form, with
little automated support. In this paper, we exploit the advantages
of information visualization and present a system that aims to
detect occupational fraud in systems which involve a pair of
entities (e.g., an employee and a client) and periodic activity. The
main visualization is based on a spiral system on which the events
are drawn appropriately according to their time-stamp. Suspicious
events are considered those which appear along the same radius or on
close radii of the spiral. Before producing the visualization, the
system ranks both involved entities according to the specifications
of the internal auditor and generates a video file of the activity
such that events with strong evidence of fraud appear first in the
video. The system is also equipped with several different
visualizations and mechanisms in order to meet the requirements of
an internal fraud detection system.
\end{abstract}


% \fi
\IEEEpeerreviewmaketitle

%-------------------------------------------------------------------------
\section{Introduction}
\label{sec:introduction}
%-------------------------------------------------------------------------

Occupational fraud represents a serious and continuous threat for
companies worldwide regardless of their size or type, and may cause
severe damage to the operation of a company. Occupational or
employee fraud can be defined as \emph{the intentionally misuse or
abuse of the resources of a company by an employee that takes
advantage of the employment position for personal profit}. Fraud
cases that are considered as occupational fraud include the
following: (i)~falsification of financial statements, (ii)~asset
misappropriation, and (iii)~bribery or corruption of employees.

According to a recent survey of the Association of Certified Fraud
Examiners~\cite{ACFE12}, fraud causes a $5\%$ loss to companies'
revenues each year, which applied to the estimated $2011$ Gross
World Product, leads to a potential global loss of more than $\$3.5$
trillion, while the median loss caused by the occupational fraud in
the survey was $\$140.000$. Apart from the economic loss, a company
that is victim of fraud has to face many other non-negligible
consequences. Among them is the loss of reputation towards its
customers, employees, and other entities (financial institutions,
vendors, etc.), especially in cases where the companies keep record
of personal data and/or transactions.

For these reasons, the prevention and detection of fraud within a
company is of tremendous importance, but it remains a problem of
outstanding difficulty and requires severe internal control.
However, monitoring of companies' anti-fraud control systems is a
time-consuming task that requires huge effort since the log data
generated by these systems are in textual form and their amount is
not easily manageable by the internal auditors in daily basis not
even in a weekly basis. Thus, the detection of malicious events and
the corresponding response to them cannot be immediate.

Examining occupational fraud schemes in specific systems in which an
employee and a client are involved (e.g., billing, membership
renewal systems, etc) reveals that events that occur in regular time
basis may be indications of fraud. For instance, in a billing system
of a company, if a specific employee appears to have a monthly
activity towards an account of a customer, this should be considered
as a suspicious periodic series of events that has to be further
examined. These events become more important in the case they occur
before the billing date of the client or outside the employee's
working hours.

\begin{figure*}[h!tb]
  \centering
  \includegraphics[width=\textwidth]{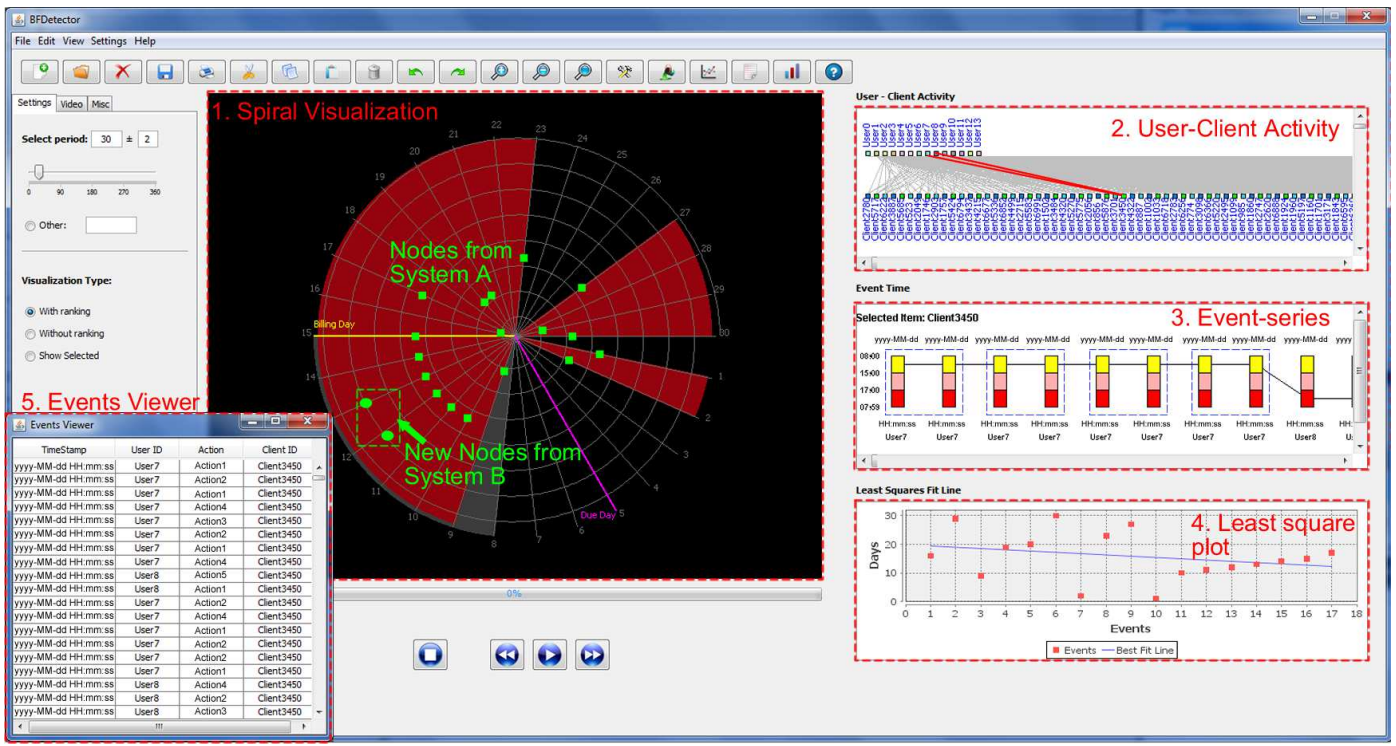}
  \caption{A snapshot of the interface of the system. Dates, usernames and actions are made anonymous for confidentiality reasons.}
  \label{fig:snapshot}
\end{figure*}

In this paper, we present a system that visualizes serial data
produced by business control systems in which a pair of entities
(e.g., employee-client) is involved. The main goal of our system is
to detect periodic patterns suggesting that an employee possibly
falsifies the invoices and/or the account of a client.
Figure~\ref{fig:snapshot} illustrates a snapshot of our system. The
base of our system is a spiral visualization on which the time-stamp
of each event is appropriately represented. The main advantage of
spiral visualizations is that potential periodic patterns can be
quickly identified since they appear along a radius or on close
radii of the spiral. Our system consists of several coordinated
visualization windows, each dedicated to a particular aspect of
audit data. The top-rightmost visualization of
Figure~\ref{fig:snapshot} illustrates the total activity of
employees and clients. The middle panel at the right side of
Figure~\ref{fig:snapshot} demonstrates the time and the dates for
each event related to a client using different colors to identify
access during specific time intervals of a day. This visualization
contributes to the quick identification of events that appear
outside the employees' working hours or on holidays, which may be
indications of fraud. The bottom-rightmost visualization is a least
square plot where the $y$-axis corresponds to the days of a month
and which aims to detect periodicity. In the case where the plotted
line is ``almost'' parallel to $x$-axis and the points are ``close''
to this line, this implies that there exists a periodic pattern
related to a day of the month. There also exits an event-viewer
(refer to panel $\#5$ of Figure~\ref{fig:snapshot}) that illustrates
the initial input data and interacts with the visualization.

The proposed system has been developed based on feedback provided by
internal auditors a major Greek company. The main obstacle reported
while trying to detect occupational fraud is the amount of data that
is usually generated from more than one business systems and has to
be investigated manually by writing and executing scripts. For this
reason, we have tried to develop a system that quickly detects
periodic patterns in data and incorporates several of the common
patterns that are investigated by the auditors in order to detect
occupational fraud. In addition, since occupational fraud is a
sensitive issue, auditors have also emphasized the necessity to have
a tool that is able to quickly confirm or reject their suspicions
about the activity of an employee. The proposed system can also be
used for this purpose.

The system is user-oriented and the visualization can be adapted
appropriately such that it depicts the patterns that are
investigated by the auditor. The innovation of this tool is that it
aggregates the total activity of each employee and client and ranks
both of them according to specifications defined by the internal
auditor. Based on this ranking the system produces a video file.
Frames are ordered such that those with strong evidence of fraud
appear first in the video.

The system is also equipped with supplementary visualizations that
provide information about the activity of the employees and clients.
In order to meet the requirements of an internal auditor, the system
supports supplementary functionalities such as filtering, export log
mechanisms, storing, reloading and post-processing of data. It
provides also advanced graphic functionality, including popup menus,
printing capabilities, custom zoom, fit-in window, selection,
dragging and resizing of objects.

This paper is structured as follows: Section~\ref{sec:overview}
overviews the detection procedure. In Section~\ref{sec:ranking}, we
present the factors based on which the system ranks the employees
and the clients. In Section~\ref{sec:description}, we describe in
detail the system and the visualization features. In
Section~\ref{sec:case-study}, we present a case-study on real data
from a Greek company. We conclude in Section~\ref{sec:conclusions}
with open problems and future work.

% ============================================================================
\section{Related Work}
\label{sec:previous-work}
% ============================================================================

Over the last few years, much research effort has been focused on
the field of fraud detection and several diverse approaches have
been proposed. However, to the best of our knowledge, there exist
only few papers that deal exclusively with occupational fraud
detection. For this reason, in this section, we mention research
that tries to detect various types of fraud. Most of the existing
work makes use of data-mining techniques. An overview of existing
publications on data-mining can be found in \cite{BH02, KLSH04,
Lue10, PKSG10}. There exist also, several approaches that present
relational data-mining techniques using the graph structure that may
be applied to fraud detection. Among them, Eberle and
Holder~\cite{EH09} presented a graph-based anomaly detection
approach in order to detect occupational fraud in business
transactions and processes. In their work, they search for anomalous
instances of structural patterns which are hidden in data that
represent entities, relationships and actions.

Pattern matching and graph-pattern matching approaches have also
been proposed for fraud detection and several systems have been
developed. We will name only a few. The NASD Regulation Detection
System (ADS)~\cite{KSH*S98, Sen00, SGS*02} monitors trades and
quotations in the Nasdaq stock market and tries to detect patterns
and practices of violative activity. The Financial Crimes
Enforcement Network AI System (FAIS)~\cite{GS95, SGW*95} is a system
that detects money laundering in cash transaction data. The Link
Analysis Workbench (LAW)~\cite{WBH*03} is developed to detect cases
of terrorist and other criminal activity in noisy and incomplete
information. Luell~\cite{Lue10} presented a system that combines
data-mining and graph-pattern matching techniques in order to detect
occupational fraud within a financial institution. Visual data
mining has also, been applied for fraud detection. Huang et
al.~\cite{HLN09} presented a framework of visual analytics for stock
market security using $3D$-treemaps. Chang et al.~\cite{CLG*08}
presented ``WireVis'', a system that uses interactive visualization
techniques in order to search for suspicious financial transactions.
Di~Giacomo et al.~\cite{GDLP10} proposed a system based on
information visualization techniques to discover financial crimes.
Didimo et al.~\cite{DLMP11} developed a system that supports the
analyst with effective tools in order to discover financial crimes,
like money laundering and frauds. Didimo et al.~\cite{DLM12}
presented VIS4AUI, a system that collects financial information with
regard to ongoing bank relationships and high value transactions and
tries to detect money laundering cases. Stasko et al.~\cite{SGL08}
developed a visual analytic system that facilitates analysts to
examine reports and documents more efficiently in order to identify
potential embedded threats.

Regarding the identification of periodic patterns in serial data,
Carlis and Konstan~\cite{CK98} suggested an approach in which serial
attributes of data are represented along a spiral axis, while
periodic ones along the radii of the spiral. Weber et
al.~\cite{WMM01} presented an approach which uses spirals to
visualize large sets of time-series data and reveals periodic
structures. In their approach, the time axis is represented by the
spiral and other features of the data are depicted by points,
colors, bars or lines. Bertini et al.~\cite{BHL07} presented
SpiralView, a tool that visualizes on a spiral the distribution of
network alarms over time and helps revealing potential periodic
patterns. Argyriou and Symvonis~\cite{AS12} suggested visualization
techniques based on concentric circles which aim to quickly identify
periodic events in serial data in order to reveal occupational
fraud. Suntinger et al.~\cite{SOSG08} developed a visualization
system that represents events from event-based systems on a
cylindrical tunnel. The visualization detects several incidents,
such as particular patterns or irregularities that might affect
business performance. Regarding the visualization of time-series
data and the available techniques, an overview can be found in
\cite{ABMTS07, MS03, SC00}.

As mentioned above, our system was developed under the guidance of
internal auditors of a major Greek company and it is designed to
meet their requirements during the investigation of fraud cases.
Their major requirement was to design a system that reveals
reoccurring activity between pairs of employee-clients. For this
reason, we have adopted the spiral visualization that enables serial
data visualization. However, in contrast with the work of Bertini et
al.~\cite{BHL07} that used the spiral in order to visualize events
daily, the data-sets that we had to process cover a much larger time
interval and had to be visualized simultaneously. This fact combined
with the large volume of the data-set had to be faced efficiently in
order to avoid cluttering the visualization. Also, unlike to the
data-sets used for financial crime detection where there exists a
``hint'' on suspicious cases or transactions, in our data-sets, in
most of the cases, it is highly unlikely to have prior knowledge on
suspicious cases. Hence, it is difficult and maybe ``risky'' to try
to filter the data before producing the visualization. This
motivated us to design the ranking procedure in order to distinguish
suspicious cases that have to be further examined and try to adapt
it to the needs of the auditors. If compared with other existing
systems for financial crime detection or employee fraud mentioned
above, our system (i)~produces a video file containing activity of
clients-employees according to the time interval selected by the
auditors, (ii)~distinguishes suspicious events based on the ranking
factors selected by the auditor and distributes them in the video
frames such that they are quickly identified, and (iii)~supports
multiple coordinated views that facilitate the investigation and
reveal the periodic activity.

% ============================================================================
\section{Overview of the Detection Procedure}
\label{sec:overview}
% ============================================================================

In this section, we present an overview of the fraud detection
procedure as supported by our system. The input data consist of log
files or sets of database records generated by systems which involve
pairs of entities (e.g., billing systems). Each record may have been
generated by a call between an employee and a client, a transaction
involving both entities, etc. Hence, a record consists of a
time-stamp, an employee, a client and an action taken by the
employee. Since the log files of a company are usually generated by
different control systems that support different log mechanisms
(e.g., databases, files, etc), the input data are appropriately
parsed and stored in the database of the system. Then, the system
ranks the clients and the employees according to rules specified by
the auditor and creates a video with all the activity of a client.
By default, the ranking is calculated based on the entire database,
unless the auditor specifies a desired time-interval. In the
visualization, all potentially suspicious incidents are detected and
can be further investigated by the auditor, who makes use of the
system's accompanied visualizations. Before we proceed with the
detailed description of the system, we introduce some terminology
necessary for the description of the ranking procedure.

An \emph{event} $e$ involves a pair of employee-client and is
defined by a $4$-tuple $(t,u,c,a)$, where:
\begin{itemize}
  \item $t$ is the time-stamp of the occurrence of the event,
  \item $u$ is the id of the employee,
  \item $c$ is the id of the client,
  \item $a$ is the action taken by the employee.
\end{itemize}
For a particular $4$-tuple $(t,u,c,a)$, we say that client $c$
\emph{is related to} event $e$ and is also \emph{related to}
employee $u$. For a client $c$, an \emph{event-series} $T_c =
\{e_1^c, e_2^c, \ldots, \}$ is a sequence of events
$e_i^c=(t_i,u_i,c,a_i)$ related to client $c$. Note that such an
event-series consists of events which may involve more than one
employee.

% ============================================================================
\section{Ranking Procedure Description}
\label{sec:ranking}
% ============================================================================

In general, the ranking of a client is based on several factors such
as the number of events close to the billing date, the actions taken
by the employees, and so on, which are described in detail in
Section~\ref{sec:client-ranking}. The ranking of employees is based
on the ranking of the clients that are related to a specific
employee. Note that, since a company may have thousands of clients,
in the ranking procedure we take into consideration only the ones
for which there exists an event generated by a log mechanism (i.e.,
not all registered clients of the business system of a company).

% ============================================================================
\subsection{Client Ranking Function}
% ============================================================================
\label{sec:client-ranking}

In order to rank clients, the system analyzes the event-series that
correspond to each client based on factors defined by the auditor.
In the system, we have incorporated several of the queries commonly
used by internal auditors, while seeking for occupational fraud.

Let $N$ be the number of distinct factors considered in the ranking
calculation. Let also $a_f^c$ be the ``performance'' of client $c$
at factor $f$. According to the severity (low, medium, high) of the
corresponding event-series, $a_f^c$ equals to zero, one or two,
respectively. Ranking $R_c$ of client $c$ is defined as follows:
$$R_c = \sum\limits_{f=1}^Na_f^c \cdot w_f,$$ where $w_f$ is the weight of importance
for factor $f$.

The weights of factors $w_f, f=1,\ldots,N$ are determined by the
auditor, who also specifies an ordering among them that best fits to
what he/she is seeking for. For instance, if the auditor is
interested in events that occur outside the employee's working
hours, the corresponding factor should be ranked first, which
implies that the weight of the corresponding factor should be
greater than the weights of other factors supported by the system.
Given a factor ordering, weights $w_f, f=1,\ldots,N$ of the factors
are calculated based on a formula proposed by Stillwell et
al.~\cite{SSE81}, as follows:
$$w_f = \frac{N-r_f+1}{\sum\limits_{j=1}^N(N-r_j+1)},$$
where $r_f$ is the rank position of factor $f$ in the factor
ordering.

In the following, we describe the factors that are currently
supported by the system. For each of these factors, we define three
classes of clients according to the severity (low, medium, high) of
event-series $T_c$ corresponding to client $c$. Then, performance
$a_f^c$ of client $c$ on factor $f$ is defined by the following
formula:

   \begin{displaymath}
    a_f^c = \left\{
        \begin{array}{lr}
        2, & T_c \in High~Severity~Class~for~factor~f~~~~~\\
        1, & T_c \in Medium~Severity~Class~for~factor~f\\
        0, & T_c \in Low~Severity~Class~for~factor~f~~~~~~~
        \end{array}
    \right.
    \end{displaymath}

Note that the default values that define each of the above classes
and are described in the remainder of this section were suggested by
the auditors of the company according to their requirements.
However, they can be appropriately alternated, if needed.

% ============================================================================
\subsubsection{Distance from Billing Date}
% ============================================================================

Experience on examining occupational fraud schemes has shown that
events related to the same pair of employee-client that appear on a
monthly basis and before the billing date of a client's invoice may
be strong indications of fraud. Given an event-series $T_c$
corresponding to client $c$, the system detects events whose
time-stamp is close to the billing date of client $c$. Based on the
number of such events, the system calculates the severity of
event-series $T_c$. In the case where the time-stamp of an event
occurs after the billing date of a particular month, we consider
this as an incident that concerns next month's activity.

Let $e$ be an arbitrary event and let $t_e$ be its time-stamp. Let
also $t_e'$ be the billing date that immediately follows $t_e$.
Then, the distance of event $e$ from the billing date, denoted by
$d_e$, is defined by the number of days between $t_e$ and $t_e'$.

Let $D_c^0$ be the set of events which occur within distance of
three days from the billing date, i.e., $D_c^0 = \{e^c \in T_c : d_e
\leq 3 \}$ and $|D_c^0|$ its cardinality. Similarly, we define the
set of events which occur within distance greater than three and
less than seven days from the billing date, i.e., $D_c^1 = \{e^c \in
T_c: 3 < d_e \leq 7 \}$ and the set of events $D_c^2 = \{e^c \in T_c
: d_e > 7 \}$ with distance more than seven days from the billing
date. The end-points of the above investigated time-intervals can be
adjusted if desired, by the auditor. The evaluation of the
importance of this factor is based on a classification of
event-series $T_c$ in one of the following severity classes
according to the number of events occurred close to the billing
date. More precisely, event-series $T_c$ related to client $c$,
belongs to this class if:

\emph{High Severity Class:} This class includes event-series with
severe indications of fraud for which it holds one of the following:
  \begin{itemize}
    \item $|D_c^0| \geq 2$: This implies that there
    exist at least two events related to client $c$ too close to the
    billing date. To minimize false-positives, cases where there exists only one event too close to
    the billing date are not considered by the system of high-severity, since they
    may have occurred by coincidence. However, they are classified
    to a medium-severity class in order to be further investigated.
    \item $|D_c^1| \geq 3$: We included this case in high severity class,
    since the specific event-series contains an ``unusual'' number of events within distance of one week from the billing date.
    Again, the system tries to minimize false-positives by taking into consideration events
    that occurred within the interval of $(3,7]$ days from the billing date at least three times.
    \item $|D_c^0| + |D_c^1| \geq 2$: In this case, the system takes into consideration the total number
    of events occurred within distance of one week from the billing
    date.
  \end{itemize}
\emph{ Medium Severity Class:} In this class, we consider
event-series for which it holds one of the following:
  \begin{itemize}
    \item $|D_c^0| = 1$ or $|D_c^1| = 1$ or $|D_c^1| = 2$: These cases were excluded from the
    high-severity class in order to minimize false-positives. However, they
    have to be investigated since these may imply that malicious activity has just begun.
    \item $|D_c^2| > thres$, where $thres$ is a threshold defined by the auditor. By default, this value is $5$. This implies that there exists a continuous activity
    concerning client $c$, which may have to be further investigated.
  \end{itemize}
\emph{Low Severity Class:} All other cases.

Similarly, one can define a factor regarding the due date of the
invoice of a client. Again, in this case we are interested in events
that occur before the due date. However, it is recommended that the
auditor does not use simultaneously the ``distance from billing
date'' factor and the ``distance from due date'' factor, since there
exist overlaps between the investigated intervals which may create
false-positives.

% ============================================================================
\subsubsection{Event-series periodicity}
\label{subsec:periodicity}
% ============================================================================

Given an event-series $T_c$ related to client $c$, we define the
\emph{proper period of activity} as its period when $T_c$ is treated
as a time-series. Then, assuming that $T_c$ is ordered according to
the time-stamps of its events, we calculate its proper period of
activity based on the algorithm presented in~\cite{AS12}. Since
internal auditors are interested in events that appear on monthly
basis, the system evaluates the severity of the event-series as
follows:

\emph{High Severity Class:} Event-series with period $p$ such that
of $27 \leq p \leq 30$ or $31$ days.

\emph{Medium Severity Class:} Event-series with period $p$ such that
of $20 \leq p < 27$ days.

\emph{Low Severity Class:} All other cases.

As previously, the auditor can adjust the values that determine the
intervals for each of the above classes.

% ============================================================================
\subsubsection{Events Occurring Outside Working Hours}
% ============================================================================

Given an event-series related to a particular client, the system
tries to detect suspicious cases occurred within a day. Events of
high-severity are the ones that occur outside the employee's working
hours, on weekends or holidays. Furthermore, events of
medium-severity are considered those that occur at the end of the
employee's shift. We have assumed that this case corresponds to the
last two hours of employee's shift. However, this value can be
appropriately adjusted by the auditor. The system takes as an input
the working hours of each employee and also takes into consideration
weekends and holidays. The classification of a client based on the
time-stamps of the related events is performed as follows:

\emph{High Severity Class: }There exists at least one event
  occurred outside working hours, on weekends or holidays.
\emph{Medium Severity Class: }There exists at least two events at
  the end of employees' shift. As previously, the system requires
  at least two such events to include an event-series in this class in order to minimize false-positives.
\emph{Low Severity Class:} All other cases.

Again, the auditor can appropriately adjust the values that define
the above classes.

% ============================================================================
\subsubsection{Number of Employees related to a Client}
% ============================================================================

This factor indicates the number of employees that are related to a
specific client. Normally, it is expected that distinct events are
related to several distinct employees. Due to the fact that, usually
a randomly selected employee handles a client request, having the
same employee handling multiple requests of the same client may be
an indication of fraud. The classification of the event-series
related to client $c$ based on this factor, is the following:

\emph{High Severity Class:} One employee handles more than $50\%$ of
the events related to client $c$.

\emph{ Medium Severity Class: }Two or three employees handle more
than $50\%$ of the events related to client $c$.

\emph{Low Severity Class:} All other cases.

The auditor can also adjust the percentages that define the above
classes. However, there may exist cases where for instance, a client
calls the company for an issue regarding his/her account and always
asks for the same employee. This obviously, does not consist fraud
and the system would provide a false-positive if this factor was the
only factor applied for ranking. Hence, it is recommended to be
applied in conjunction with other factors.

% ============================================================================
\subsubsection{Action Name}
% ============================================================================

Each company has its own rules regarding the employees that use the
business systems and supports different privileges for different
employees. Hence, there exists a list of actions that may be
forbidden for all or for unauthorized employees. In addition, there
exist actions which are suspicious, even though an employee may be
authorized to perform. Furthermore, there exist actions that may be
correlated (e.g., open-close action) and it is uncommon if one of
them does not appear in the event-series. Thus, since there exist
several rules in order to detect fraud in diverse business systems,
the auditor in our implementation has to adjust the rules of each
severity class on the corresponding panel of the system. An overview
of the classification based on this factor is the following:

\emph{High Severity Class:} Actions that are forbidden for
unauthorized employees.

\emph{Medium Severity Class:} Actions that are considered to be
  suspicious, as described above.

\emph{Low Severity Class: }All other actions.

% ============================================================================
\subsubsection{Client Status}
\label{sec:previous-client-ranking}
% ============================================================================

When ranking a client, it is important to take into consideration
its corresponding background history. This implies that a client for
which there existed evidence of fraud will be ranked higher. The
auditor is able to mark a client as (i)~blacklisted, if a previous
investigation led to evidence of fraud, (ii)~suspect, if a previous
investigation is ongoing or unresolved, or (iii)~cleared, if
suspicions of fraud either do not existed in the system or were not
confirmed. According to this marking, we define the following
classification:

\emph{High Severity Class:} The client is blacklisted.

\emph{Medium Severity Class:} The client is a suspect.

\emph{Low Severity Class:} The client is cleared.

% ============================================================================
\subsection{Employee Ranking}
% ============================================================================

As mentioned in Section~\ref{sec:overview}, the system ranks the
employees based on several aspects of their relation with their
clients. Consider an arbitrary employee $u$  and let $S_u$ be the
set of clients that are served by employee $u$, i.e., $S_u = \{c :
\exists~event~e=(t,u,c,a)\}$. Let also $R_u = \{R_c \ : c \in S_u\}$
be the set containing the rankings of these clients. By default, the
system assigns to the employee the value that corresponds to the
maximum ranking of set $R_u$. This implies that if an employee is
related to a ``suspicious'' client, then he/she will be also
considered as ``suspicious''. An alternative could be to consider
the clients with rankings greater than a threshold defined by the
auditor.

%% ============================================================================
\section{System Description}
\label{sec:description}
%% ============================================================================

In this section we describe in detail our fraud detection system.
The system operates in two modes, either \emph{off-line} or
\emph{semi-online}. In brief, in the off-line mode the system parses
static data concerning a period of time (e.g., the data of a week)
and provides corresponding visualizations. The semi-online mode can
be used on a daily basis in order to visualize the daily activity of
the employees and clients. In both modes, the system provides
interactive visualizations that help in the detection of suspicious
events. Visualizations of large data-sets may not be useful in
certain cases. To cope with this problem, the auditor is able to
specify a time-window and then, the system visualizes events whose
time-stamp belongs to the query window. However, ranking can be
estimated either based on the whole data-set that includes data from
a much longer  period of time or on data occurring in the specific
time window, according to the specifications of the auditor.

\subsection{Off-line mode}

As mentioned in Section~\ref{sec:introduction}, the system consists
of multiple coordinated views and each of them visualizes a
different aspect of the audit-data.

\subsubsection{The spiral visualization}

\begin{figure}[h!tb]
  \centering
  \includegraphics[width=0.5\textwidth]{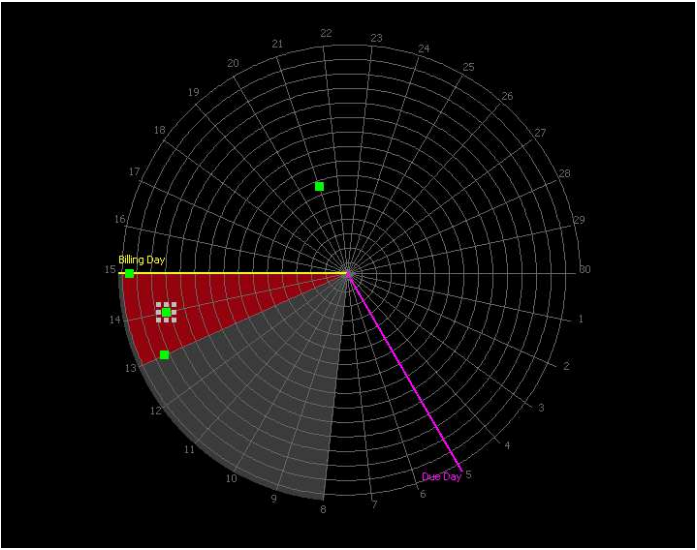}
  \caption{The main visualization when the system operates in off-line mode.}
  \label{fig:system-interface}
\end{figure}

A snapshot of the system in off-line mode is illustrated in
Figure~\ref{fig:system-interface}. In the spiral visualization, each
spiral branch visualizes a period of one month, while the number of
spiral branches is related to the first and last time-stamp of the
input data (if not alternatively selected by the auditor), starting
from the first month that coincides with the inner branch of the
spiral. Each spiral branch is split by a number of lines according
to the periodicity value that is examined (i.e., $7$ days, $15$
days, $30$ days, etc.) and each line corresponds to a day of a
month. The default value is $30$ days, which implies that the
administrator seeks for monthly suspicious activity (see
Figure~\ref{fig:system-interface}).

In the spiral window we place nodes, where each node represents an
event related to a client and its position is determined based on
the corresponding time-stamp. Nodes of different colors represent
events related to different clients. To produce the spiral
visualization, the system ignores multiple appearances of events
that correspond to the same pair of employee-client at the same
date. According to the spiral structure, events related to the same
client and appearing along a radius of a spiral are considered
suspicious and need to be further examined. However, examining cases
of fraud has revealed that suspicious events may not always appear
on the same date from month to month, and thus, suspicious events
may appear on close radii. These should also be considered as
suspicious.

Employees and clients will be ranked according to the specifications
of the auditor and the ranking function mentioned in
Section~\ref{sec:ranking}. Based on the ranking, the system
generates a video file in which each frame depicts the activity of a
client within the specified time interval, giving priority to the
ones with the higher ranking.  The ordering of the frames guarantees
that clients that are considered to be suspicious will not be
skipped during processing; even in cases of large data sets they
will be immediately distinguished.

The auditor is able to pause the video in order to further
investigate the activity of a client. By default, on each frame the
billing and the due date of the corresponding client are depicted on
the visualization (see Figure~\ref{fig:system-interface}). The
light-gray colored region of Figure~\ref{fig:system-interface}
corresponds to the ``dangerous'' interval of a week before the
billing date. The red-colored region of
Figure~\ref{fig:system-interface} indicates that there exist events
from month to month that differ by less than $3$ days. If necessary,
the auditor is able to visualize nodes that involve the same
employee by the same color. By these features, the auditor quickly
identifies events that occur close to the billing/due date and
potential periodic patterns for a specific client.

Filtering techniques are also supported by the system. The ranking
factors described in Section~\ref{sec:ranking} can also be used as
filters while the results can be exported in separate log files. For
instance, from a single frame, the auditor can select only the nodes
representing events occurred outside the employee's working hours.
The auditor can also perform custom queries to the database which is
a fundamental functionality for fraud detection. Optionally, the
system is able to save a produced visualization in a file for the
case where post-processing is required. It also maintains records
about the employees/clients activities and their ranking.

As already mentioned, by default, the ordering of the frames is
based on the ranking assignment. However, the nodes of the
visualization can be distributed on the frames according to an
ordering specified by the auditor, which may be based on predefined
knowledge about a client or on a list of clients already marked by
the auditor from a previous investigation.

\subsubsection{Supplementary visualizations}

\begin{figure}[h!tb]
  \includegraphics[width=0.5\textwidth]{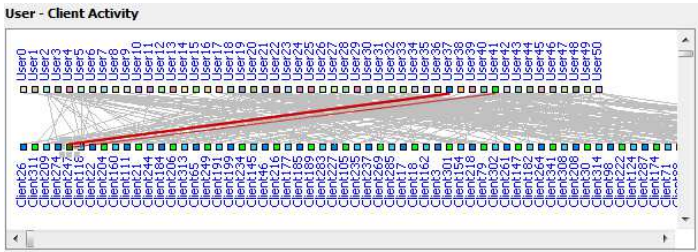}
  \caption{A $2$-layered visualization representing the total activity of employees and clients.}
  \label{fig:layered}
\end{figure}

We proceed to describe the supplementary visualizations of the
system. Figure~\ref{fig:layered} depicts a $2$-layer visualization
representing the total employee-client activity of the input data.
The upper layer corresponds to distinct employees, while the bottom
one to distinct clients. The ordering of the clients at the bottom
layer is according to their ranking assignment. The node coloring
follows the one used for each client in the spiral visualization.
This visualization contributes to quickly identify pairs of
employees-clients that appear to be involved in many events and
simultaneously gives an overview of the total activity of the
entities. Also, it demonstrates employees involved with many
``suspicious'' clients. The visualization interacts with the spiral
drawing such that when a pair of employee-client is selected from
the spiral drawing, it is also, marked in the layered visualization,
and vice-versa. In the case where there exist more that one event
related to the same pair of employee-client the thickness of the
corresponding edge becomes larger. Optionally, the visualization can
be filtered such that only the client that is displayed in the video
along with its related employees is visualized.

\begin{figure}[h!tb]
  \includegraphics[width=0.5\textwidth]{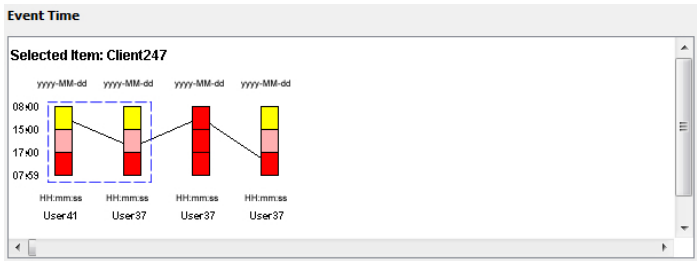}
  \caption{A visualization that depicts an event-series related to a client and distinguishes the time that an event occurs within a day. Dates are represented by yyyy-MM-dd for confidentiality reasons.}
  \label{fig:line-visualization}
\end{figure}

In the visualization depicted in
Figure~\ref{fig:line-visualization}, the event-series related to a
specified client is represented by a line-graph. Each node of the
drawing corresponds to the day of occurrence of an event. Each such
node is split into time intervals that correspond to the hours of
the day. The middle part (refer to the pink-colored region of
Figure~\ref{fig:line-visualization}) corresponds to the end of the
shift of the employee (i.e., two last hours of the shift). The upper
part (refer to the yellow-colored region of
Figure~\ref{fig:line-visualization}) corresponds to the employee's
time-shift having excluded the last two hours of the shift, while
the bottom part (refer to red-colored regions of
Figure~\ref{fig:line-visualization}) to non-working hours of a day.
The endpoints of an edge touch the parts that correspond to the
time-stamp of the events. Note that for the spiral visualization the
system ignores multiple events that correspond to the same pair of
employee-client. For the visualization of
Figure~\ref{fig:line-visualization}, in the case where multiple
events for the same pair of employee-client occur within a day,
multiple nodes will be drawn and will be bounded by a rectangle in
order to be distinguished. The system optionally takes as an input
the shifts for each employee and makes the proper adjustments to the
visualization. Also, since weekends and holidays can be taken under
consideration by the system, if such cases occur the corresponding
nodes are entirely colored in red (refer to the red-colored node of
Figure~\ref{fig:line-visualization}).

\begin{figure}[h!tb]
  \includegraphics[width=0.5\textwidth]{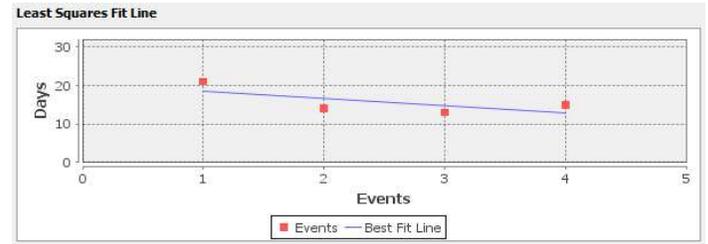}
  \caption{A least-square plot that indicates whether there exist periodic events within a time interval.}
  \label{fig:least-squares}
\end{figure}

Given the event-series of a client, the system provides a plot where
each point $(x,y)$ represents the day $y$ of the period interval
that event $x$ occurred (see Figure~\ref{fig:least-squares}). Then,
using the least squares method~\cite{DS81}, the line that best fits
to the data-set is calculated and plotted. Cases where the slope of
the line tends to zero (i.e, almost parallel to $x$-axis) and points
are close to the line (the model fits well to the point-set)
indicate that most of the events appear close to the same day of the
month. This implies that there exists a ``suspicious'' periodic
pattern (e.g., close to day $15$). Studying the least-square plot,
we have to take into consideration cases where the calculated line
is ``almost'' parallel to $x$-axis but, the points are not close to
it. In this case, the impression of the periodicity is fictionally,
since the line does not fit well to the point-set. However, these
cases can be distinguished quickly either visually or by taking into
consideration the least-square model error.

Note that, all the above visualizations are updated while the video
frames change. This ensures that the auditor has a full view of the
activity of each client without changing screens or drawbacks. A
supplementary panel, as the one at the left-most part of
Figure~\ref{fig:snapshot}, demonstrates the database records related
to the client in textual form such that the auditor does not have to
recall database records in order to see the initial input. The panel
also, interacts with the spiral visualization such that when
selecting a node from the spiral the corresponding event in the
panel is selected and vice-versa.

\begin{figure}[h!tb]
    \centering
  \includegraphics[width=0.4\textwidth]{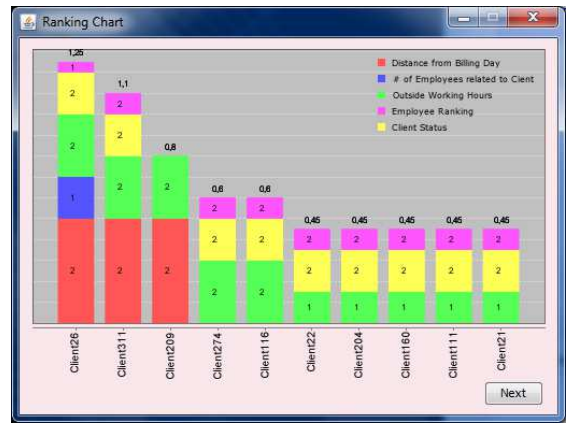}
  \caption{Client ranking based on $5$ factors. Ten most highly-ranked clients are presented.}
  \label{fig:ranking}
\end{figure}

Another feature of the system is that it provides a stacked bar plot
demonstrating the ranking of the clients as illustrated in
Figure~\ref{fig:ranking}. Each bar corresponds to a client and is
split into regions that represent each of the factors used for
client ranking, while their lengths are proportional to the
performance of the client on this factor. Also, in each of these
regions the performance of the client on this factor (i.e., 0, 1, or
2) is illustrated. The plot of Figure~\ref{fig:ranking},
demonstrates the ten highly-ranked clients based on five factors.
The first bar indicates that Client-$26$ was in the High-Severity
Class (refer to Section~\ref{sec:ranking}) to three out of the five
factors calculated, since client's performance was $2$ on Factors
$1$, $3$ and $4$.

\begin{figure}[h!tb]
  \centering
  \includegraphics[width=0.4\textwidth]{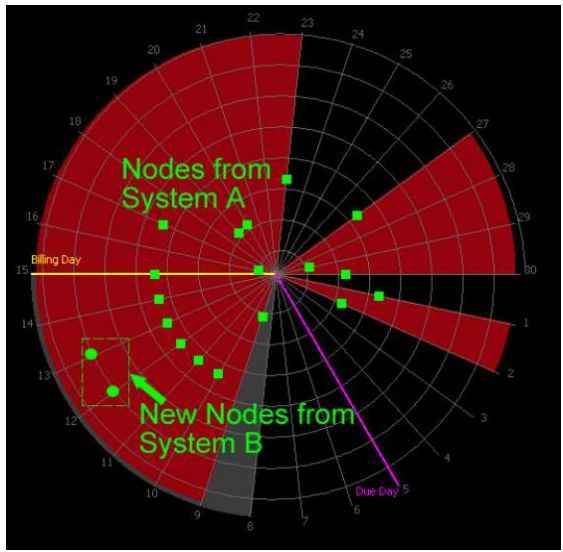}
  \caption{Different shapes correspond to events generated from different business systems.
   Their combination in the visualization reveals the periodic activity and probably fraud.}
  \label{fig:different-systems}
\end{figure}

Feedback provided by internal auditors added to our system another
important functionality when searching for fraud. Sometimes, in
order to detect fraud it is necessary to trace suspicious events to
more than one business systems. For this reason, the auditor is able
to select a client and load its activity from more than one such
systems. In this case, the nodes of the visualization are drawn with
different shape. In the visualization of
Figure~\ref{fig:different-systems}, rectangular nodes correspond to
a business system, say A. In the case where circular nodes
corresponding to a different system, say B, are loaded to the
visualization, our suspicions on periodic patterns are much more
justified.

\begin{figure}[h!tb]
  \centering
  \includegraphics[width=0.5\textwidth]{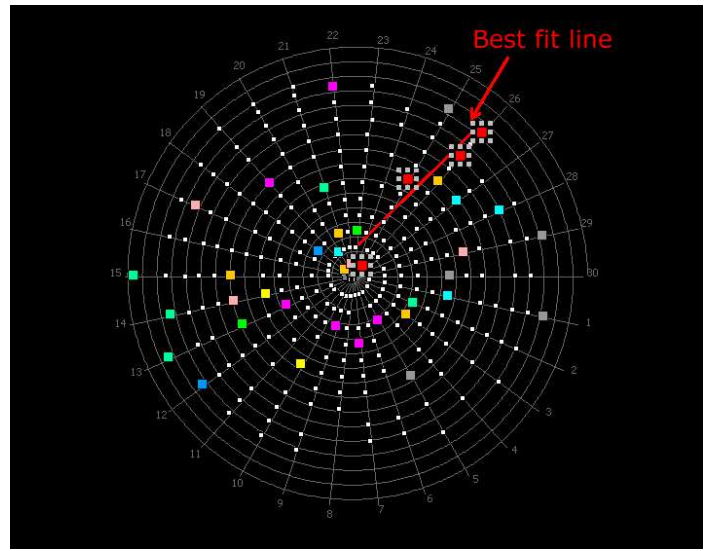}
  \caption{Spiral visualization when data are visualized without filtering.}
  \label{fig:unfiltered-view}
\end{figure}

The system also provides an unfiltered view of the processed data,
as illustrated in Figure~\ref{fig:unfiltered-view}. Colored nodes
correspond to ten highly-ranked clients. Nodes that correspond to
the same client will be represented by the same color. However,
since it is difficult to distinguish suspicious behavior in such
visualizations the system optionally draws a line that best fits to
the activity of the selected client. Filtering techniques like the
ones mentioned above are also supported. By default, the
visualization is filtered such that clients related to only one
event are excluded.

\subsection{Semi-online mode}

When developing the system we have tried to incorporate
functionalities for processing dynamic data. However, it would be
impossible in real conditions to continuously have an auditor
monitoring the company's activity. Feedback provided by internal
auditors also discouraged us, since fraud analysis takes place on a
system different than the operational system which generates log
records. For these reasons, we have developed functionalities for
processing the data daily as soon as they are generated by the
systems.

\begin{figure}[h!tb]
  \centering
  \includegraphics[width=0.5\textwidth]{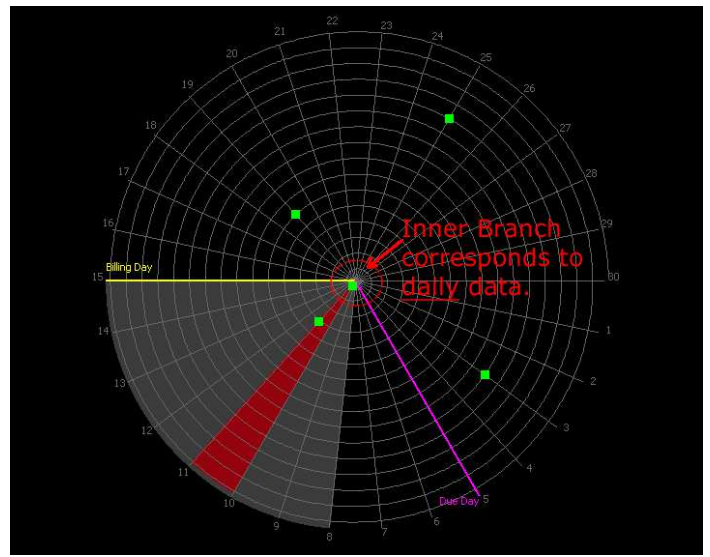}
  \caption{Spiral visualization when the system operates at semi-online mode.}
  \label{fig:semi-online}
\end{figure}

The main visualization of the semi-online mode is again a spiral and
all visualization features are similar to the corresponding ones of
the off-line mode (refer to Figure~\ref{fig:semi-online}). In
contrast to the off-line, in semi-online mode, the inner branch of
the spiral corresponds to the events of current day. The other
branches coincide to the first and last month of the input data. The
spiral is split by lines representing the days of a month. In order
to produce the visualization, the system re-ranks both the clients
and employees based on the whole data-set (i.e., including both
previous and new records) and defines their ordering in the video
frames. In semi-online mode, each client related to an event is
visualized at the appropriate position according to its time-stamp
on the inner branch along with all previous events related to that
client. Again, the auditor seeks for events that appear along a
radius or on close radii of the spiral. This visualization
facilitates the immediate detection of a periodic pattern that may
have just begun.

% ============================================================================
\section{Case Study}
\label{sec:case-study}
% ============================================================================

In this section, we present a case study on real data from a major
Greek company (the company name cannot be revealed due to a
nondisclosure agreement). For confidentiality reasons, the data
presented in this case study were preprocessed and made anonymous.
The data-set which we processed corresponds to a time interval of
six months and consists of approximately $35.000$ entries involving
$7200$ distinct clients and $14$ employees stemming from a single
fraud management system of the company. Note that the data from most
of the fraud management systems of the company include sensitive
personal data and we were not allowed to process them for the
purposes of our case study. The case study was performed while the
system operated in off-line mode and without taking into
consideration any prior client or employee ranking.

Since the case study was conducted in collaboration with the fraud
experts of the company, the main question raised by them was to
identify pairs of employee-client that appear to have more than ten
events during the last six months, i.e., the time-interval of the
data-set. In order to become familiar with the data-set, we first
ranked the clients based on the number of their events. The system
identified $430$ clients whose number of related events is larger
than ten (about $6\%$ of the number of clients in the initial
data-set) and distributed them accordingly in the video frames. Even
though this number is much smaller with respect to the total number
of clients in the data-set, the investigation was still a hard task
for the auditor (due to the number of clients to be examined). In
the next step, we performed a second ranking (as described in
Section~\ref{sec:ranking}) on the clients based on three factors:
(i)~the number of events related to a client, (ii)~the number of
actions that are highly unlikely to appear and are indications of
possible fraudulent activity, and (iii)~the number of distinct
employees serving the client. Precise details on the configuration
values of each ranking factor are omitted due to a nondisclosure
agreement. Also, since we were not communicated the information
about the billing date of each client and the employees' shifts we
had to ignore these ranking factors.

The ranking procedure distinguished $52$ out of $430$ clients (about
$0.7\%$ of the number of clients in the initial data-set) that were
highly-ranked in the above factors and presented them in the first
frames of the video. In the next frames, the system presented $62$
clients (about $0.9\%$ of the number of clients in the initial
data-set) that were medium-ranked in the above factors. The results
were further investigated by the auditors who tried to detect
periodic patterns (daily or monthly) in the specific frames. The
auditors also suggested to apply the periodicity factor described in
Section~\ref{subsec:periodicity} in order to detect events that
appear within a time interval of (i)~$28$ days and (ii)~$5$ days.
For the first case, the system identified $3$ out of $52$ clients
that were highly-ranked and $6$ out of $62$ clients that were
medium-ranked. For the second case, $11$ out of $52$ highly-ranked
clients and $12$ out of $62$ medium-ranked clients were identified.
For the final step of the investigation, the auditors used the
supplementary visualizations and the log viewer of the system along
with their experience and supplementary data that were not
communicated to us in order to evaluate the severity of these
events. It should be emphasized that the real-time investigation of
the data performed by the auditors (i.e., the non-automated log file
processing) did not identify any of these reoccurring activity. The
results were taken into consideration by the auditors for further
internal investigation.

% ============================================================================
\section{Conclusions and Future Work}
\label{sec:conclusions}
% ============================================================================

This paper proposes a system that aims to detect occupational fraud
in business systems in which a pair of entities, such as
employee-client are involved (e.g., billings systems, membership
renewal system, etc). The system operates in off-line and
semi-online mode. The main visualization consists of a spiral axis
on which the data are mapped to a specific position according to the
time they occur. Periodic events that appear to a radius of the
spiral or on radii close to each other, are suspicious and need to
be further examined. Our work is on-going and opens several
directions for future work:

\begin{itemize}
\item Regarding the ranking function, more factors have to
be taken into consideration in order to produce a more accurate
value.

\item It would be better for the auditor to add custom factors and define the appropriate functions through the
graphical user interface of the system.

\item Incorporating more functionality that may be useful for an
administrator such as statistic analysis of the activity for each
entity, plots, bar charts, etc.

\item It will be of interest to search for groups of collaborating employees/clients that
may be involved in suspicious events, performing different
clustering techniques.
\end{itemize}
%
% ============================================================================
\section*{Acknowledgements}
% ============================================================================

The work of Evmorfia N. Argyriou has been co-financed by the
European Union (European Social Fund - ESF) and Greek national funds
through the Operational Program "Education and Lifelong Learning" of
the National Strategic Reference Framework (NSRF) - Research Funding
Program: Heracleitus II. Investing in knowledge society through the
European Social Fund.



%\bibliographystyle{IEEEtran}
%\begin{thebibliography}{1}
%\end{thebibliography}

\bibliographystyle{IEEEtran}
\bibliography{references}

\end{document}